\documentclass[aps, prd, amsmath, floats,floatfix, twocolumn,superscriptaddress, nofootinbib]{revtex4}
\usepackage{amsmath} 
\usepackage{longtable}
\usepackage{verbatim} 
\usepackage{mathrsfs}
\usepackage{amsfonts}
\usepackage{dsfont}
\usepackage{color}
\usepackage{graphicx}
\usepackage{tikz}

\newcommand{\be}{\begin{equation}}
\newcommand{\ee}{\end{equation}}
\newcommand{\bea}{\begin{eqnarray}}
\newcommand{\eea}{\end{eqnarray}}

\begin{document}

\title{Occam’s Razor and a Three-Family Ultralight Fermionic Dark Matter Model}
\author{Fabian Hernandez-Gutierrez}
\email{f.hernandezgutierrez@ugto.mx}
\affiliation{Division de Ciencias e Ingenier\'ias,  Universidad de Guanajuato, Campus Leon,
 C.P. 37150, Le\'on, Guanajuato, M\'exico.}
\author{J. Barranco}
\email{jbarranc@fisica.ugto.mx}
\affiliation{Division de Ciencias e Ingenier\'ias,  Universidad de Guanajuato, Campus Leon,
 C.P. 37150, Le\'on, Guanajuato, M\'exico.}

\date{\today}

\begin{abstract}

Inspired by the three generations in the Standard Model, we propose that dark matter particles are simply three non-interacting ultralight fermions. This model is tested at galactic scales with the SPARC catalog of rotational velocity curves. This is done by modeling the dark matter halo as the self-gravitating configuration
of three gravitationally coupled fermion species. This configuration is capable of forming more complex rotational curves which gives it flexibility to describe more complex and diverse morphologies of galaxy rotation curves. The core of the halos will be described by degenerate fermions and the transitions from a degenerate core to a non-degenerate halo will be described by either a perfect-fluid-like halo or a NFW density profile. 
In order to construct the priors of the three-fermion model, first
we performed a single-fermion analysis and obtained three characteristic mass regions around 20, 40, and 65-70 eV, for both halo types.
It is found that the three-component fermion model improves the goodness of fit for 94.9\% of the galaxies considered. To estimate if the additional parameters are favored despite the added complexity by the corresponding improvement in the fit, the information criteria of AICc and BIC were computed and we found that more than $50\%$ of the galaxies analyzed  prefer the three-component fermion model over a single fermion model. 
The three-fermion model is particularly favored  for galaxies with larger radial extent and more complex rotation curves. The multicomponent fermionic dark sector is able to describe a large diversity of rotation curves while maintaining characteristic fermion mass scales across the galaxy sample.
\end{abstract}
\maketitle
\section{Introduction}
The quest for the fundamental building blocks of nature is a history developed during years of experiments and observational empirical evidence that has reached the astrophysical era. 
The first elementary particle, the electron, was first detected in 1897 by J. J. Thomson, who demonstrated that cathode rays consisted of universal, negatively charged particles \cite{Thomson1897}. 
Forty years later, Anderson and Neddermeyer found the muon, now considered a fundamental charged particle \cite{Neddermeyer1937}. Although similar to the electron, the trajectory and penetrating power of this new particle in the cloud chamber were consistent with a particle 207 times heavier than the electron. The discovery of the muon provided the first indication of heavier replicas of particles with otherwise the same properties. A third particle, the tau, was later discovered \cite{Perl:1975bf} by studying anomalous electron-positron annihilation events that can be explained by a particle with the same quantum numbers as the electron but 3447 times heavier than the electron.
These heavier counterparts are known as generations, and three generations exist for each type of Standard Model fermions, which we consider fundamental building blocks of nature.
Within each generation, particles share the same fundamental interactions and quantum numbers, while differing in their masses. 
With the help of gigantic particle accelerators, the quarks, the fundamental fermions that form the nucleus of the atoms were discovered \cite{Breidenbach:1969kd,Gell-Mann:1964ewy,Zweig:1964ruk,Botton,top,D0:1995jca}.
Quarks experience strong, electromagnetic, and weak interactions; while charged leptons experience electromagnetic and weak interactions.
Finally, neutrinos were detected in 1953 \cite{Reines:1953pu}, and they again consist of a family of three particles, but in this case they interact only through weak interactions. All of them are subject to gravity. The experimental detection of neutrinos involved the construction of massive detectors due to their feeble interaction with matter and in 2003, it was confirmed that they are three generations of massive particles  \cite{Super-Kamiokande:1998kpq,SNO:2001kpb}.

Then we look at the sky.
Fritz Zwicky’s 1933 observation of the Coma Cluster provided an early indication of a discrepancy between luminous matter and dynamical observations \cite{Zwicky1933}. 

Several decades later, Vera Rubin reported a systematic study of galactic rotation curves: 21 Sc galaxies encompassing a wide range of radii, masses, and luminosities \cite{Rubin1980} exhibited the need for additional non-luminous mass, showing that this was not an isolated case, but a general issue in astrophysical phenomena. 

After this seminal work, advances in astronomical observations, cosmology, and numerical simulations established dark matter as a central question in physics, and led to the development of the Cold Dark Matter (CDM) paradigm \cite{1982ApJ...263L...1P,Blumenthal:1984bp,Davis:1985rj,bardeen1986statistics,Efstathiou:1990xe,COBE:1992syq,WMAP:2003elm,Planck2018}.
The empirical evidence, either by direct detection experiments \cite{PandaX-4T:2021bab,LZ:2022lsv,XENON:2023cxc} or by indirect searches \cite{Barman:2022jdg,Chan:2019ptd,Song:2023xdk} of dark matter, suggests that it is matter that interacts only through gravity.
This structure motivates a dark fermionic sector composed of three distinct particles with hierarchical masses that are only coupled to ordinary matter through gravity. This hierarchy is schematically illustrated in Fig. \ref{figure:Standard_Model_particles}. 
\usetikzlibrary{decorations.pathreplacing}

\newsavebox{\tempbox}

\tikzstyle{particleblue}=[
    rectangle,
    thick,
    draw=black,
    fill=blue!20,
    minimum size=25pt
]

\tikzstyle{particleyellow}=[
    rectangle,
    thick,
    draw=black,
    fill=yellow!40,
    minimum size=25pt
]

\tikzstyle{particleorange}=[
    rectangle,
    thick,
    draw=black,
    fill=orange!40,
    minimum size=25pt
]

\tikzstyle{generation}=[
    rectangle,
    draw=white,
    fill=white
]

\newcommand\particleblue[2]{%
    \node (#1) [particleblue] {${\boldsymbol{#2}}$};
}

\newcommand\particleyellow[2]{%
    \node (#1) [particleyellow] {${\boldsymbol{#2}}$};
}

\newcommand\particleorange[2]{%
    \node (#1) [particleorange] {${\boldsymbol{#2}}$};
}

\newcommand\generation[2]{%
    \node (#1) [generation] {\tiny${\boldsymbol{#2}}$};
} 

\begin{figure}[htbp]
\centering

\begin{lrbox}{\tempbox}
\begin{tikzpicture}[>=latex,text depth=0.00ex]

    \matrix[row sep=0.05cm,column sep=0.05cm,nodes={align=left}]{
        \generation{I}{I}&
        \generation{II}{II}&
        \generation{III}{III}&
        \\

        \particleblue{u}{u}&
        \particleblue{c}{c}&
        \particleblue{t}{t}&
        \\

        \particleblue{d}{d}&
        \particleblue{s}{s}&
        \particleblue{b}{b}&
        \\[0.15cm]

        \particleyellow{e}{e}&
        \particleyellow{mu}{\mu}&
        \particleyellow{tau}{\tau}&
        \\[0.15cm]

        \particleyellow{nu_e}{\nu_e}&
        \particleyellow{nu_mu}{\nu_\mu}&
        \particleyellow{nu_tau}{\nu_\tau}&
        \\[0.15cm]

        \particleorange{f_1}{f_1}&
        \particleorange{f_2}{f_2}&
        \particleorange{f_3}{f_3}&
        \\
    };

    \draw[decorate,decoration={brace,amplitude=4pt}]
        ([xshift=4pt]t.north east) --
        ([xshift=4pt]b.south east)
        node[midway,right=8pt,align=left]
        {Strong, weak, EM\\and gravitational};

    \draw[decorate,decoration={brace,amplitude=4pt}]
        ([xshift=4pt]tau.north east) --
        ([xshift=4pt]tau.south east)
        node[midway,right=8pt,align=left]
        {Weak, EM\\and gravitational};

    \draw[decorate,decoration={brace,amplitude=4pt}]
        ([xshift=4pt]nu_tau.north east) --
        ([xshift=4pt]nu_tau.south east)
        node[midway,right=8pt,align=left]
        {Weak and gravitational};

    \draw[decorate,decoration={brace,amplitude=4pt}]
        ([xshift=4pt]f_3.north east) --
        ([xshift=4pt]f_3.south east)
        node[midway,right=8pt,align=left]
        {Gravitational};

\end{tikzpicture}
\end{lrbox}

\resizebox{0.5\textwidth}{!}{\usebox{\tempbox}}

\caption{A three-fermion dark matter extension of the Standard Model. Quarks are shown in blue, leptons in yellow, and the additional dark-sector fermions in orange.}
\label{figure:Standard_Model_particles}

\end{figure}
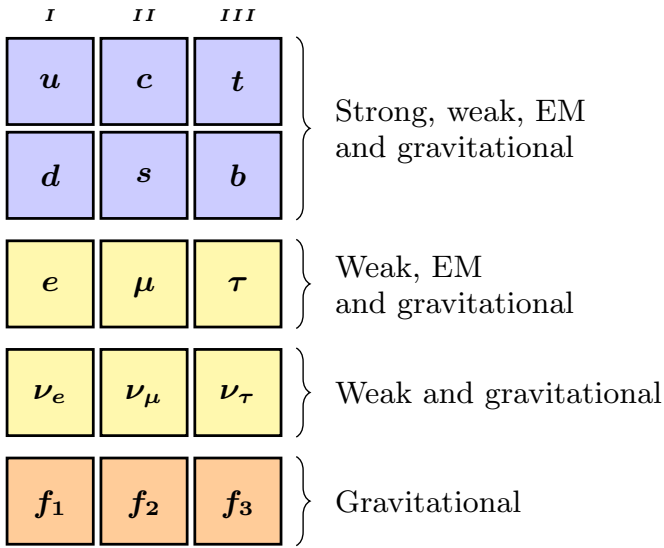

For most cosmological data the paradigm of CDM is suitable, as it reproduces large-scale phenomena efficiently \cite{Planck2018}. It is at galactic and sub-galactic scales where CDM remains in tension with observations. 

Simulations based on the CDM framework result in the well-known Navarro-Frenk-White (NFW) dark matter profile \cite{1997Navarro}, which predicts a steep density profile as it nears the center of the halo; this is in disagreement with observations and is known as the Cusp-core problem \cite{Bullock2017}. Other small-scale problems include: the Missing Satellite problem \cite{Klypin1999}, the number of predicted satellite galaxies is larger than observed, and the Too-Big-to-Fail problem \cite{Boylan-kolchin2011}, the average density of dwarf galaxies is larger than observed in the local group.

The study of galactic-scale phenomena, such as  rotation curves and velocity dispersions, is of much importance. The diversity of galaxy types and morphologies suggests that dark matter halos could have more complex structures than the one predicted by the standard NFW profile \cite{Oman:2015xda,Zentner:2022xux,Sands:2024vvd}. 

The structure and configuration of the resulting dark matter halo arise from the properties of the dark matter candidate. Early studies of dense stellar matter showed that the Pauli exclusion principle, through degeneracy pressure, plays a crucial role in supporting white dwarfs against gravitational collapse \cite{Fowler1926}. This motivated the hypothesis that dark matter could be made of fermionic particles, in which case the same degeneracy pressure would support the dark matter halo.

The possibility of quantum effects on galactic scales implies a minimum radius below which a purely classical description ceases to be valid \cite{Destri2013}. This would turn the self-gravitating system into a natural quantum macroscopic object, which would make it necessary to consider quantum mechanical calculations to compute galaxy structures at kpc scales and below, which the classical N-body simulations do not explicitly account for. For the innermost structure of the halo, where quantum mechanical effects are essential, classical $N$-body simulations would not be appropriate. This could help explain some of the tensions between $N$-body simulations and the observed dark matter halos.

In this scenario, multiple studies have shown that a sufficiently light fermionic particle can form a self-gravitating system whose central regions are supported by the Pauli exclusion principle \cite{Destri2013, Domcke2015, Randall2017, Chavanis2015}. These keV or sub-keV particles can produce a cored density profile which could solve the \textit{core-cusp} problem. 
Nevertheless, a single fermion cannot explain the diversity of galaxies, an issue that can be circumvented by the introduction of self-interaction \cite{Hernandez-Gutierrez:2025zhg,Kamada:2016euw}.
Furthermore, to address the diversity problem, the study of self-gravitating objects has included more complex, multi-component systems such as fermion-fermion \cite{Jin1999, Jin2000, Lopes2026} stars, as well as $\ell$-boson stars \cite{NavarroBoullosa2023}. Other composite systems include fermion-boson stars \cite{Henriques1989, Henriques1990, Jetzer1990}  and fermion-soliton stars \cite{LeePang1987, DelGrosso2023}. Note that in the case of boson stars, the cores are supported by the gradient energy
of the scalar field and not by the degeneracy pressure, as is the case for fermions.

Among the possible multi-fermion configurations we believe one stands out naturally: the case of three fermion species.

It is difficult to implement the production of fermionic dark matter in the early Universe \cite{Dodelson:1993je,Shi:1998km,Shaposhnikov:2008pf}. In this work we will not address  the production mechanism, but since we are considering fermions completely decoupled from the standard model interactions, a non-thermal production mechanism is more plausible \cite{Baer:2014eja}. 

Some recent works have shown that gravitational production of singlet fermions is possible in other formulations of general relativity \cite{Barman:2019mlj,Rigouzzo:2023sbb}. These type of models can produce the observed abundance of dark matter in a wide range of fermion masses. 
Instead, in this work we will simply assume the existence of three dark matter fermions and we hope that this work motivates a further cosmological analysis. 

In this work we develop a self-gravitating system of three fermions, with distinct masses and judge the cost-benefit of this model against the single fermion model. The model consists of three gravitationally coupled fermion species, which each have a fully degenerate and non-degenerate regime. Our approach is different from the case of sterile neutrinos \cite{Dodelson:1993je,Shi:1998km,Shaposhnikov:2008pf} since we are not trying to explain or solve other problems of the standard model of particles.

The existence of three fermion species makes the halo model more flexible, but results in a model with four more free parameters to determine. This naturally raises the question of whether the improvement in the description of galaxy rotation curves is sufficient to justify the added complexity; Occam's razor implies that a simpler model is preferable. One must assess whether the goodness of fit is such that the added parameters could be justified through model-selection criteria.

In Sec. \ref{sec:fully-degenerate} we introduce the multi-component fermionic self-gravitating model, which is modeled through a generalized form of the Tolman-Oppenheimer-Volkoff (TOV) system, for the case where the system consists of three fully degenerate components, the transition from a degenerate core to a non-degenerate halo for each component is developed in Sec. \ref{sec:semi-degenerate}. To identify possible characteristic mass regions for each fermion in Sec. \ref{sec:mass-regions} we analyze the SPARC galaxies \cite{Lelli_2016} within the single-fermion scenario and using a Gaussian mixture model the resulting distribution of the posterior median fermion masses obtained is decomposed into three Gaussian distributions. Sec. \ref{sec:three-fermion-bayes} presents the Bayesian analysis of the three-fermion model using the previous one fermion analysis as the basis for the priors. Both models are compared using the Akaike and Bayesian information criteria in Sec. \ref{sec:model-comparison}. Finally, Sec. \ref{sec:conclusion} summarizes our conclusions.

\section{Three fermionic self-gravitating system: fully degenerate case}\label{sec:fully-degenerate}

After the development of Fermi-Dirac statistics and the understanding of the importance of the Pauli exclusion principle in stellar matter, Chandrasekhar developed the description of a self-gravitating system of fully degenerate fermions \cite{Chandrasekhar1939}. In the degenerate limit, the equations that describe the energy density $\rho$ and pressure $p$ are 

\begin{equation}
     p=\frac{m^{4}}{24\pi^{2}}\left[(2z^{3}-3z)\left(1+z^{2}\right)^{1/2}+3\sinh^{-1}{z}\right]
\end{equation}

\begin{equation}
    \rho=\frac{m^{4}}{8\pi^{2}}\left[\left(2z^{3}+z\right)\sqrt{1+z^{2}}-\sinh^{-1}{z}\right],
\end{equation}

which depend on $z=k_{F}/m_f$, the dimensionless Fermi momentum, where $k_F$ is the Fermi momentum and $m_f$ is the mass of the fermion.

Taking the non-relativistic limit where $z \ll 1$, that is when the mass of the fermion is much greater than the Fermi momentum $m_{f} \gg k_{F}$ , we find the equation of state for a free gas of fermions in the non-relativistic case:

\begin{equation}\label{eos_fermion}
    p=\frac{(3\pi^{2})^{2/3}}{5}\frac{\rho^{5/3}}{m_{f}^{8/3}},
\end{equation}

which corresponds to a polytrope with $\gamma=5/3$.

For a multicomponent system  of three fermions, the stress-energy tensor, $T^{\mu\nu}$, is the sum of three stress-energy tensors, each behaving in the form of a perfect fluid.

\begin{equation}
    T^{\mu\nu}=\sum_{i}^{3}  T_{(i)}^{\mu\nu}
\end{equation}

The assumption that the fermions only interact gravitationally implies the covariant conservation equations

\begin{equation}
    T_{(i)}^{\nu\mu}\ _{;\nu}=0
\end{equation}

We assume that they form spherically symmetric self-gravitating systems. The solution for a static, spherically symmetric system is given by the Tolman-Oppenheimer-Volkoff (TOV) equations.

\begin{equation}
    \frac{dp_{i}}{dr}=-\frac{GM}{r^{2}}\rho_{i}\left(1+\frac{p_{i}}{\rho_{i}}\right)\left(\frac{4\pi Pr^{3}}{M}+1\right)\left(1-\frac{2MG}{r}\right)^{-1},
\end{equation}

with $P=p_{1} + p_{2} + p_{3}$. To close the system, our equation of state is given by Eq. \ref{eos_fermion}. Considering low compactness, that is $GM<<1$ and $p_{i}<<\rho_{i}$, this simplifies the system of equations to

\begin{equation}
     \frac{dM}{dr}=4\pi r^{2} \left(\rho_{1}+\rho_{2}+\rho_{3}\right)
\end{equation}

\begin{equation}
    \frac{dp_{i}}{dr}=-\frac{GM}{r^{2}}\rho_{i}
\end{equation}

\begin{equation}
    \frac{d\rho_{i}}{dr}=-\left(\frac{3}{\pi^{2}}\right)^{1/3}m_{f}^{8/3}\frac{GM}{r^{2}}\rho_{i}^{1/3}
\end{equation}

These equations can be solved numerically, with boundary conditions $M(r=0)=0$ and $\rho_{i}(r=0)=\rho_{0,i}$, which is the central density of the $i$-th fermion and a free parameter to be fitted. The system has in total six free parameters, the masses and central densities of the fermions.\\

The system consists of three concentric spheres, each composed of one fermion species and a finite radius that fulfill the condition $\rho_{i}(R_{i})=0$, where $R_{i}$ is the radius of the spherically symmetric volume occupied by the $i$-th gas.

\begin{figure}[htbp]
    \centering
    \includegraphics[width=\linewidth]{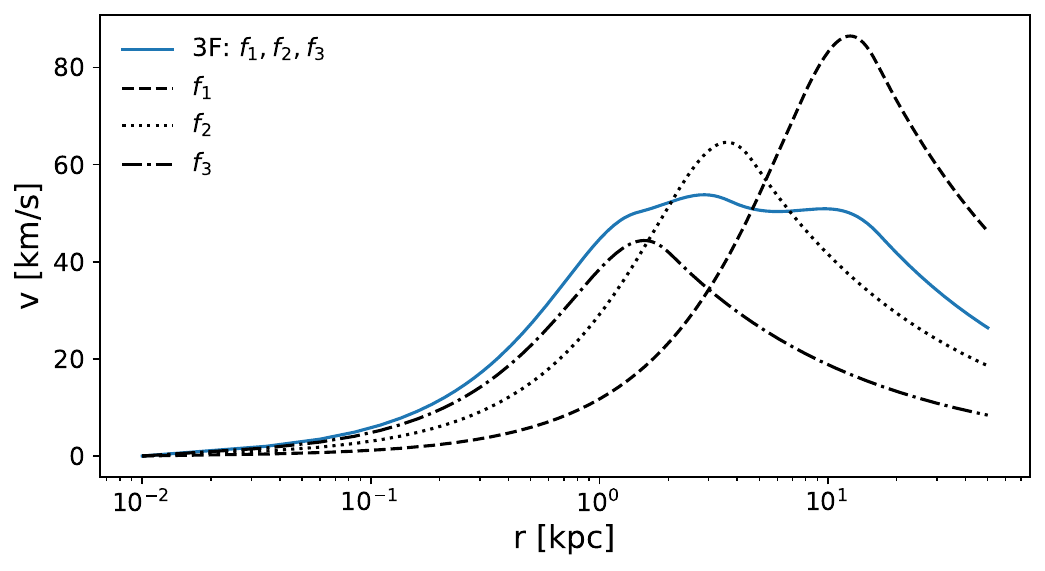}
    \caption{Rotation curve of the fully degenerate three-fermion system (blue), compared with the corresponding single-fermion cases (black). The corresponding parameters are: $\mathbf{m}_f=(30,60,100)\,\mathrm{eV}$, $\boldsymbol{\rho}_0=(0.3,2,5)\,\mathrm{GeV\,cm^{-3}}$}
    \label{fig:3F_rotcurve_degenerate}
\end{figure}

\begin{figure}[htbp]
    \centering
    \includegraphics[width=\linewidth]{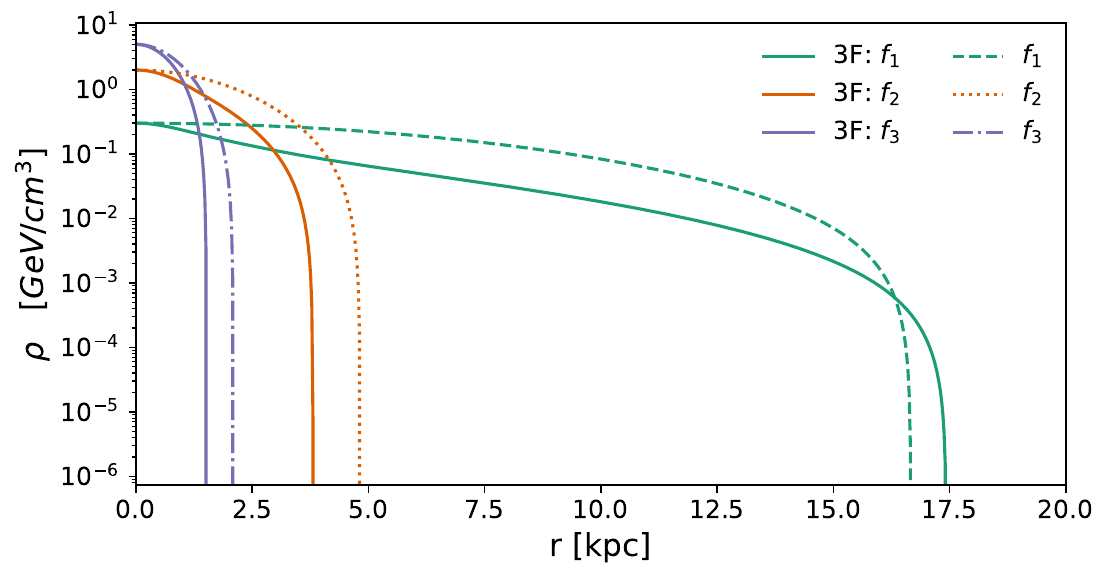}
    \caption{Density profiles of the three fermionic components in the fully degenerate three-fluid system (solid), compared with the independent single-fermion solutions (dashed).
    $\mathbf{m}_f=(30,60,100)\,\mathrm{eV}$, $\boldsymbol{\rho}_{0}=(0.3,2,5)\,\mathrm{GeV\,cm^{-3}}$}
    \label{fig:3F_density_degenerate}
\end{figure}

The self-gravitating system has six different free parameters: the mass of each fermionic species $m_{f,i}$, and the central density of each fermionic gas component $\rho_{0,i}$. These parameters influence the size and mass of the system, creating a variety of possible structures.

The presence of the other gravitationally coupled species of fermions modifies the density profile relative to the corresponding single-fermion case, as seen in Fig. \ref{fig:3F_density_degenerate}. In the single fermion case the mass of the fermion influences the radius of the self-gravitating system, with lighter fermion masses leading to extended configurations, while heavier fermions form more compact ones. When the other fermions are present, the size and shape of each component changes. The components closer to the center become more compact, while the outermost becomes more extended, but its density decreases more rapidly. The rotation curve of the three-fermion system reflects these changes in the density profiles, as seen in Fig. \ref{fig:3F_rotcurve_degenerate}, and shows that the rotation curve of the fully degenerate three-fermion configuration is not a simple superposition of the single-fermion systems.

\section{Three fermionic self-gravitating system: semi-degenerate case}\label{sec:semi-degenerate}

The fully degenerate description of the system  is only valid as long as the temperature of the system sits below the degeneracy temperature \cite{Carroll:1009754}. This needs to be evaluated for each of the components in the system. The degeneracy temperature of each fermionic component is defined as

\begin{equation}
    T<T_{deg}=\frac{1}{2 m_{f}}\left(\frac{3 \pi^2\rho(r)}{m_{f}}\right)^{2/3}.
\end{equation}

For a non-relativistic self-gravitating gas composed of different particle species, the local pressure can be expressed as the sum of the partial pressures of each component \cite{deVega2002}

\begin{equation}
    P(r)=\frac{T}{V}\sum_{i}^{N}\frac{\rho_{i}(r)}{m_{f,i}}
\end{equation}

The temperature can be estimated from the virial theorem. The total kinetic energy of the multicomponent gas is obtained by summing the contribution of each species, 

\begin{equation}
K=\frac{3}{2} T\sum_i N_i,
\end{equation}

 where $N_{i}=M_i/m_{f,i}$. This means that the effective temperature of the system can be determined at any point using the virial theorem

\begin{equation}
    T=\frac{4\pi G}{3}\left[\sum^{N}_{i=1}\frac{M_{i}}{m_{f,i}}\right]^{-1}\int_{0}^{R}M_{T}(r)\rho_{T}(r)rdr
\end{equation}

where $M_T$ is the total enclosed mass of the compact object, and $\rho_T$ is the sum of the densities of all components.

This effective temperature is used to determine when each species ceases to be degenerate. Every species can exit the degenerate regime at a different radius, each generating an independent core-shell configuration.

To describe the non-degenerate region of each component, two possibilities are chosen. We consider the traditional NFW-like halo and a perfect-fluid-like halo, which correspond to the description of a dilute fermion with an equation of state of the form $p=\omega\rho$.

\begin{figure}
    \centering
\includegraphics[width=\linewidth]{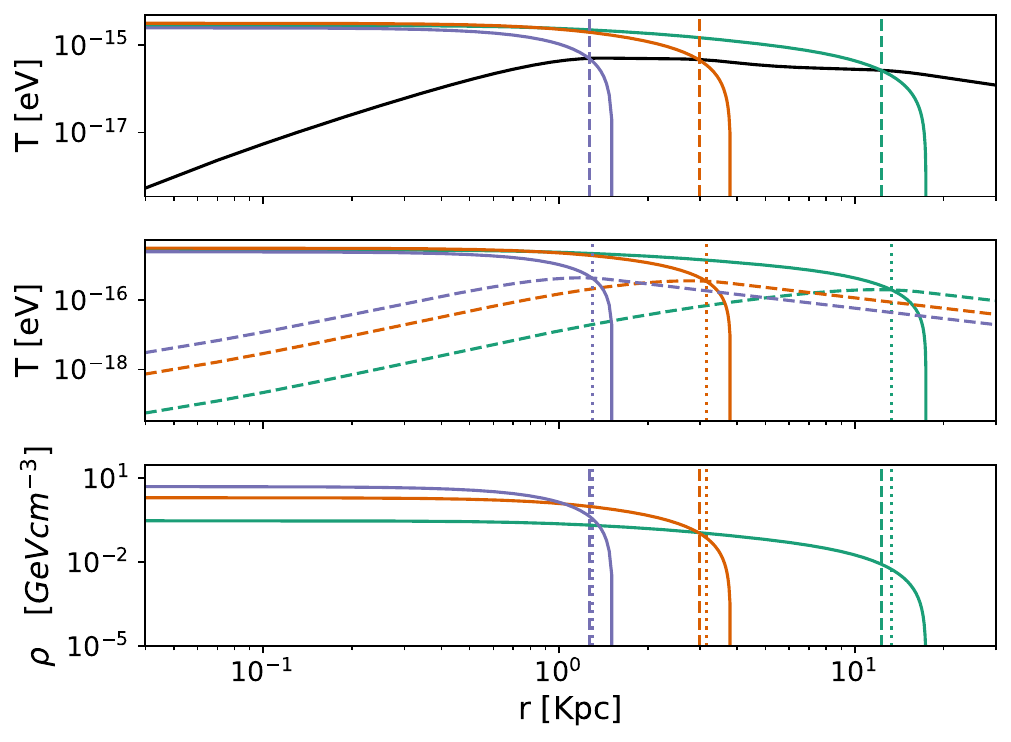}
\caption{Degeneracy transition of the three fermionic components. The upper panels compare the effective virial temperature of the multicomponent system (black) with the degeneracy temperature of each species. The middle panel show what the corresponding degeneracy temperature of each fermion (dashed) would be in the single fermion case. The lower panel shows the corresponding density profiles. Vertical lines indicate the radii at which each component leaves the degenerate regime. The adopted parameters are $\mathbf{m}_f=(30,50,70)\,\mathrm{eV}$ and $\boldsymbol{\rho}_{0}=(10,20,30)\,\mathrm{GeV\,cm^{-3}}$.}
\label{fig:transition_temperatures}
\end{figure}

\begin{figure}
\centering
\includegraphics[width=\linewidth]{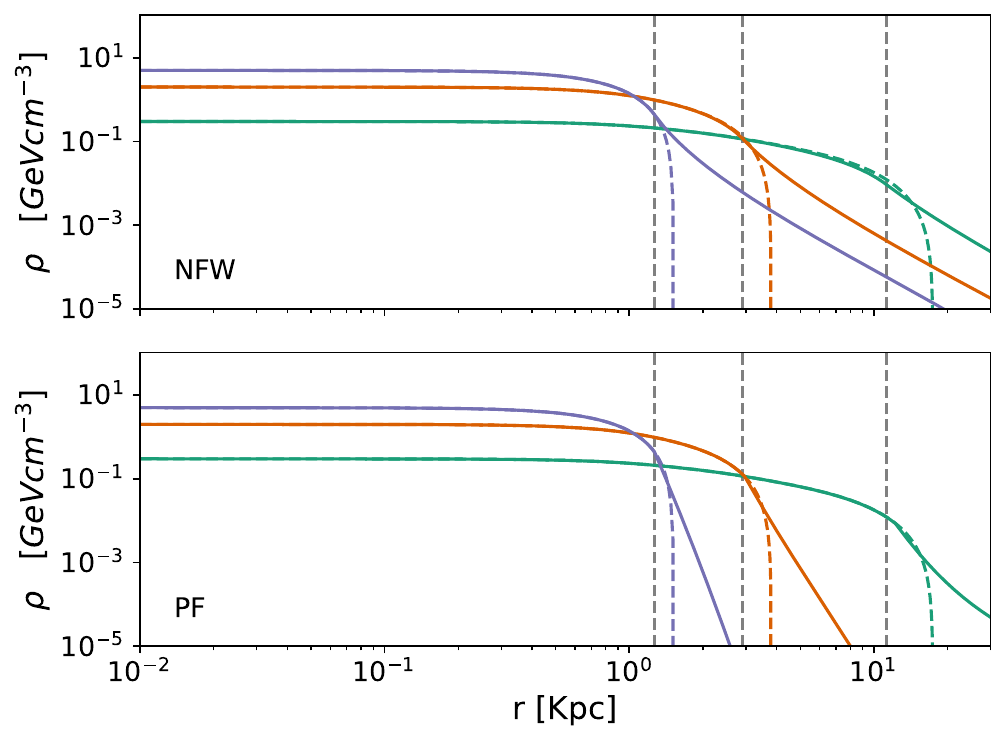}
\caption{Density profiles of the three fermionic components after the transition from the degenerate core to the non-degenerate halo. The upper panel shows the NFW case, while the lower panel shows the perfect-fluid (PF) case. Vertical lines mark the transition radius of each component. The adopted parameters are $(m_1,m_2,m_3)=(30,50,70)\,\mathrm{eV}$ and $(\rho_{0,1},\rho_{0,2},\rho_{0,3})=(10,20,30)\,\mathrm{GeV\,cm^{-3}}$.}
    \label{fig:density_transition_halos}
\end{figure}

\begin{figure}
    \centering
    \includegraphics[width=\linewidth]{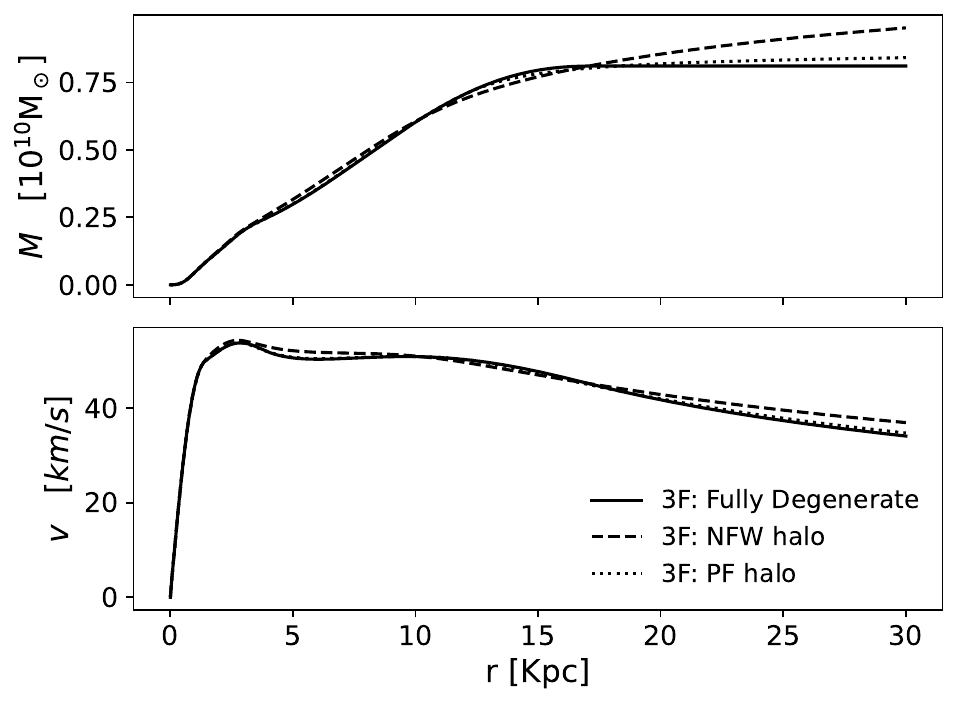}
    \caption{Comparison of the enclosed mass profile (upper panel) and rotation curve (lower panel) for the fully degenerate and both core-halo cases: NFW and PF.  The adopted parameters are $(m_1,m_2,m_3)=(30,50,70)\,\mathrm{eV}$ and $(\rho_{0,1},\rho_{0,2},\rho_{0,3})=(10,20,30)\,\mathrm{GeV\,cm^{-3}}$.}
    \label{fig:rot_curve_halo_types}
\end{figure}

For the degenerate gas of fermions to transition to a perfect-fluid-like halo, we require hydrostatic equilibrium at the radius where the fermion gas stops being degenerate and transitions. That is $p_{core}(r_{\star})=p_{halo}(r_{\star})$. This means that $\omega$ is related to $r_{\star}$ and the model parameters with the equation

\begin{equation}
    \omega=\frac{(3\pi^{2})^{2/3}}{5}\frac{\rho_{\star}^{2/3}}{m_{f}^{8/3}},
\end{equation}

where $\rho_{\star}$ is the value of the density at the radius $r_{\star}$.

The density profile of the perfect fluid halo is obtained by numerical integration, where the equation for the density of the TOV system is replaced by

\begin{equation} 
\label{tov pf}
    \frac{d\rho}{dr}=-\frac{1}{\omega}\frac{GM\rho}{r^{2}}.
\end{equation}

The other option for the non-degenerate region is given by the density profile obtained by CDM simulations, the NFW density profile. It is defined as \cite{1997Navarro}

\begin{equation}
    \rho_{NFW}(r)=\frac{\rho_{s}}{\frac{r}{R_{s}}\left(1+\frac{r}{R_{s}}\right)^{2}},
\end{equation}

where $\rho_{s}$ is the scale density and $R_{s}$ is the scale radius.

Dark matter in the CDM  is pressure-less, with $\omega=0$, and although an effective pressure is produced, we are not able to use hydrostatic equilibrium to ensure a smooth transition to the non-degenerate region.

To ensure a good transition, we demand that the density and its radial derivative be continuous at the transition radius $r_{\star}$. This results in the NFW parameters $R_{s}$ and $\rho_s$ to be given by

\begin{equation}
R_{s}=-r_{\star}\left[\frac{r_{\star}\rho'_{\star} + 3\rho_{\star}}{r_{\star}\rho'_{\star}+\rho_{\star}}\right],\quad \rho_{s}=-\frac{4(r_{\star}\rho'_{\star}\rho_{\star}^{3} +\rho_{\star}^{4})}{(r_{\star}\rho'_{\star} + 3\rho_{\star})^{3}},
\end{equation}

where $\rho_{\star}$ and $\rho'_{\star}$ are the density and its derivative in $r_{\star}$. The NFW profile then becomes 

\begin{equation} \label{NFW transition}
\rho(r)=\frac{4\rho_{\star}^{3}r_{\star}^{3}}
{r\left[r_{\star}\rho'_{\star}(r_{\star}-r)+\rho_{\star}(3r_{\star}-r)\right]^{2}}.
\end{equation}

Initially, the three fermionic components are evolved in the degenerate regime, the effective virial temperature is computed and compared with the degeneracy temperature of each component. When $T_{\rm vir}>T_{{\rm deg},i}$, 
the differential equation for that component is replaced by the corresponding non-degenerate halo type. Fig. \ref{fig:transition_temperatures} shows the effective virial temperature of the system and the degeneracy temperature of each fermionic component, with the corresponding transition points.

For the perfect-fluid case, the differential equation is replaced by Eq. \ref{tov pf} and in the NFW case, it is replaced by

\begin{equation}
\frac{d\rho}{dr}
=
-\rho(r)
\left[
\frac{1}{r}
-
\frac{2\left(r_{\star}\rho'_{\star}+\rho_{\star}\right)}
{r_{\star}\rho'_{\star}(r_{\star}-r)+\rho_{\star}(3r_{\star}-r)}
\right],
\end{equation}

where $\rho(r)$ is given by eq \ref{NFW transition}.

The transition from degenerate to non-degenerate is exemplified in the density profiles of Fig. \ref{fig:density_transition_halos}, while the corresponding mass profile and rotation curve is shown in Fig. \ref{fig:rot_curve_halo_types}.

\section{Single-fermion analysis: Obtaining possible mass regions}\label{sec:mass-regions}

To understand the capability of the three-fermion model to describe a wide range of galaxy morphologies, we use the \textit{Spitzer} Photometry and Accurate Rotation Curves (SPARC) database \cite{Lelli_2016}. This database contains a total of 175 galaxies with different morphologies, luminosities and surface brightness. This database also includes the inferred contribution of the baryonic components of the galaxy: the gas, the bulge and the disk. The theoretical rotation velocity of a test particle is given by:

\begin{equation}
v^{th}=\sqrt{V_{bar}^2+v_{DM}^2(r)}\,,
\end{equation}

the baryonic contribution corresponds to

\begin{equation}
    V_{bar}=\sqrt{|V_{gas}|V_{gas} + \Upsilon_{disk} |V_{disk}|V_{disk} + \Upsilon_{bul} |V_{bul}|V_{bul} }
\end{equation}

where $\Upsilon_{disk}$ and $\Upsilon_{bul}$ are the stellar mass-to-light ratios of disk and bulge components, respectively \cite{Lelli_2016}.

To avoid adding extra parameters to the analysis we will fix the mass-to-light ratio of the disk and bulge components of the galaxy. We fix $\Upsilon_{disk}$ to 0.5 and $\Upsilon_{bul}$ to 0.7, values that have been estimated as acceptable for most galaxies at $\mu_{3.6}$ 
\cite{Eskew_2012, Oh_2008, Meidt_2014, Schombert_2018, Schombert_McGaugh_2014, McGaugh_2016}.

In our model the three fermions have three distinct masses--one greater than the other ordered as $m_{f,1}<m_{f,2}<m_{f,3}$. To identify characteristic mass values that could be associated with these three components, we first study the mass distribution obtained in the single-fermion scenario.

We first analyze the SPARC galaxies assuming a single fermionic dark matter component. For each galaxy, the model is fitted independently for both cases of the non-degenerate halo: the perfect-fluid and NFW cases. The contribution of the dark matter halo is obtained from integrating numerically the TOV system through a Runge-Kutta method.

For each galaxy we define

\begin{equation}
    \chi^2(\boldsymbol{\theta})
    =
    \sum_{i=1}^{N}
    \frac{
    \left[
    V_{\rm obs}(r_i)-V_{\rm th}(r_i;\boldsymbol{\theta})
    \right]^2
    }{\sigma_i^2},
\end{equation}
where $V_{\rm obs}(r_i)$ and $\sigma_i$ are the observed rotational velocity and its uncertainty at the $i$-th radial point, respectively. The parameters to be estimated are $\boldsymbol{\theta}=(\rho_0,m_f)$, where $\rho_0$ is the central fermion density and $m_f$ is the fermion mass. Assuming independent Gaussian uncertainties, the likelihood is

\begin{equation}
    \ln \mathcal{L}(\boldsymbol{\theta})
    =
    -\frac{1}{2}\chi^2(\boldsymbol{\theta})
    -\frac{1}{2}\sum_{i=1}^{N}
    \ln\left(2\pi\sigma_i^2\right),
\end{equation}

\begin{figure*}
    \centering
    \includegraphics[width=\linewidth]{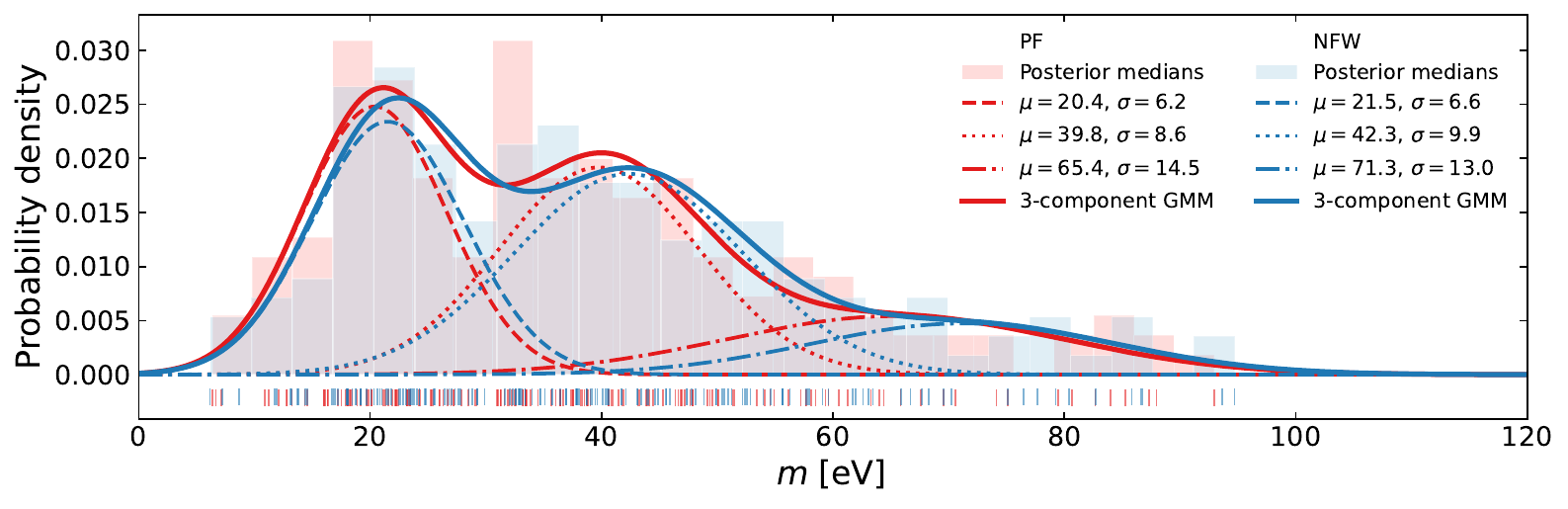}
\caption{Distribution of the posterior median fermion masses obtained from the single-fermion fits to the SPARC galaxies. The PF and NFW results are shown in red and blue, respectively. Individual line styles show the three Gaussian components of the GMM, and the solid curves show the total three-component mixture. The rug marks indicate the median mass inferred for each galaxy.}
    \label{fig:gmm_histogram}
\end{figure*}

The posterior distributions of the parameters are sampled through a Markov Chain Monte Carlo (MCMC) using the ensemble sampler implemented in \texttt{emcee} \cite{2013Foreman-Mackey}. The sampler runs 24 chains simultaneously. The initial position of the walkers is localized around the minimum-$\chi^2$ solution obtained from a preliminary grid search in $(\rho_0,m_f)$  of 900 points covering $0.01<\rho_0<80~{\rm GeV\,cm^{-3}}$ and $0.01<m_f<1000~{\rm eV}$. For both free parameters we adopt uniform priors in the ranges of $10^{-3}<\rho_0<150~{\rm GeV\,cm^{-3}}$ and $0<m_f<1500~{\rm eV}$.

The process has a 500-step warm-up phase followed by a $10^4$-step production phase and a burn-in of the first $30\%$ of the production samples. The production phase is able to stop before the $10^4$ steps are completed if a convergence criterion is satisfied. Convergence is monitored through the integrated autocorrelation time $\tau$. Every 500 steps, we require $N_{\rm step}>30\,\tau$ for all the model parameters, with a relative change in $\tau$ smaller than $5\%$.

The resulting posterior distribution of $m_f$ for each galaxy is used to characterize the possible ranges of each fermionic particle. It is evident that the distribution of median mass values is multimodal and there is no single preferred mass value.

We want to explore the possibility of dark matter being composed of three distinct fermion species, with three distinct mass values. To identify the possible mass regions, we will characterize the structure of the resulting mass distribution in terms of three Gaussian contributions. 
 
To implement this decomposition, we use a Gaussian mixture model (GMM). A GMM describes a probability distribution as a weighted sum of $K$ Gaussian components,

\begin{equation}
    p(m_f)=\sum_{j=1}^{K} w_j\, \mathcal{N}(m_f|\mu_j,\sigma_j^2),
\end{equation}
where $\mathcal{N}$ is the Gaussian distribution with mean $\mu_j$ and standard deviation $\sigma_j$ of the $j$-th component, with its relative weight  $w_j$, satisfying
\begin{equation}
    \sum_{j=1}^{K}w_j=1.
\end{equation}

We fit the distribution of posterior median masses using the GaussianMixture implementation of \texttt{scikit-learn} \cite{Pedregosa2011} with $K=3$ to obtain the characteristic mass scales of each of the three fermion species. 

To remove outliers before the GMM fit, we excluded 16 galaxies with posterior median masses lying more than 1.5 times the interquartile range beyond the central distribution.

\begin{table}[t]
\centering
\caption{Parameters obtained from the GMM fit of the distribution of posterior median fermion masses for the PF and NFW halo cases.}
\label{tab:gmm}
\begin{tabular}{ccccc}
\hline\hline
Halo & Component & $\mu$ [eV] & $\sigma$ [eV] & $w$ \\
\hline
PF  & 1 & 20.40 & 6.23 & 0.387 \\
    & 2 & 39.78 & 8.64 & 0.415 \\
    & 3 & 65.39 & 14.54 & 0.198 \\
\hline
NFW & 1 & 21.52 & 6.58 & 0.386 \\
    & 2 & 42.34 & 9.86 & 0.459 \\
    & 3 & 71.26 & 13.03 & 0.155 \\
\hline\hline
\end{tabular}
\end{table}

The results of the single-fermion posterior analysis and the GMM fit are shown in Fig. \ref{fig:gmm_histogram} and Table \ref{tab:gmm}. The mass distributions obtained are clearly multimodal, and after the GMM fit, we identify the three characteristic regions centered around 20, 40, and 65-70 eV, in both halo cases. This suggests a weak dependence on the halo type. The largest difference between the PF and NFW components appears in the third one, where the mean shifts from 65.39 eV for PF to 71.26 eV for NFW. The first two Gaussian components contain most of the total mixture weight in both cases.

\section{Three-fermion Bayesian analysis}\label{sec:three-fermion-bayes}

Motivated by the three characteristic mass scales inferred from the one-fermion analysis, we now consider a system composed of three fermionic dark matter species. The parameter space is $\boldsymbol{\theta}_{\rm 3F}=(\rho_{0,1},\rho_{0,2},\rho_{0,3},m_{f,1},m_{f,2},m_{f,3})$, with the ordering condition, where there exists a mass hierarchy that implies that $m_{f,1}<m_{f,2}<m_{f,3}$.

For the Bayesian analysis of the 3F model, we can construct data-informed priors from the mass regions obtained in the 1F analysis of the SPARC sample. For each mass parameter, we adopt a Gaussian prior with a mean and standard deviation corresponding to each component obtained from the GMM.

\begin{equation}
    \pi(m_{f,i}|\mu_i,\sigma_i^2)
    =
    \frac{1}{\sqrt{2\pi}\sigma_i}
    \exp\left[
    -\frac{(m_{f,i}-\mu_i)^2}{2\sigma_i^2}
    \right].
\end{equation}

The specific values of $\mu$ and $\sigma$ used for each component and halo type are shown in Table \ref{tab:gmm}. For the three central densities, flat priors are adopted, $0<\rho_{0,i}<150~{\rm GeV\,cm^{-3}}$.

\begin{figure*}
    \centering
    \includegraphics[width=0.8\linewidth]{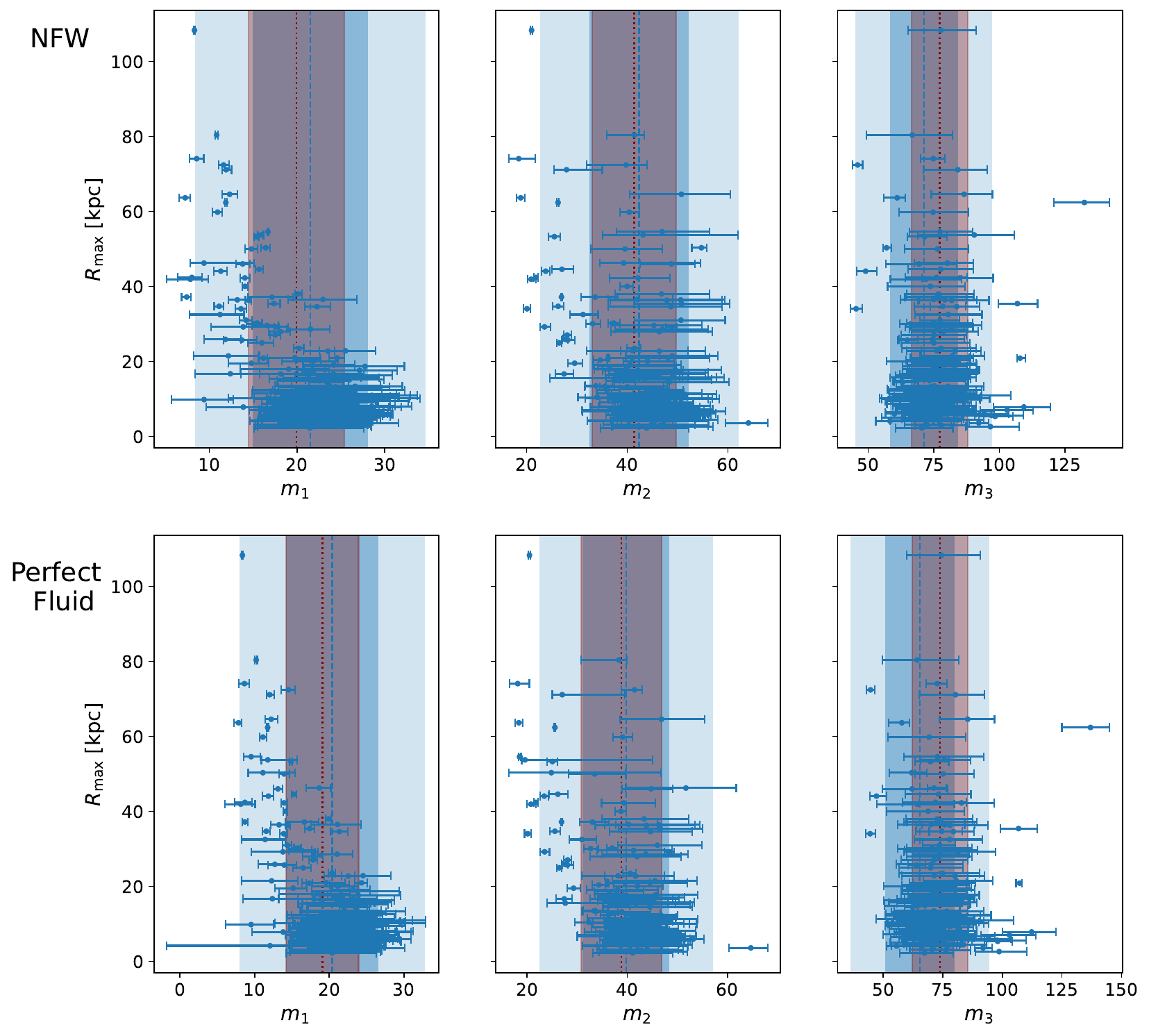}
     \caption{Posterior constraints on the three fermion masses obtained from the three-component fits to the SPARC galaxies. Each point represents the posterior median mass and a $1\sigma$ confidence interval. Blue shaded regions indicate the Gaussian priors obtained from the single-fermion GMM analysis, with the darker and lighter shaded regions corresponding to the $1\sigma$ and $2\sigma$ ranges, respectively. Maroon shaded regions show the mean and standard deviation of the posterior median masses of the three-fermion fits. Dashed vertical lines indicate the corresponding mean values.}
    \label{fig:3F_fits_radius}
\end{figure*}

As in the previous section, the posterior distributions were obtained using the \texttt{emcee} library, using MCMC with 24 walkers. 

The sampling for the three fermion case is similar to the one fermion case, with a 500 step warm-up phase and a production phase ---this time with a maximum of $2\times10^4$ steps--- and a burn-in of $30\%$. The initial position is again localized around the minimum-$\chi^2$ of a preliminary grid search of 900 points, but in this case, the masses were fixed to the means of the Gaussian components of the GMM fit.

Convergence is again monitored every 500 steps by requiring $N_{\rm step}>30\,\tau$ for every model parameter.

The characteristic mass scales inferred from the 1F analysis are largely preserved in the 3F fits, with the first two components remaining close to their prior means and the third showing a more noticeable shift toward heavier masses, corresponding to $0.46\sigma_{\rm prior}$ and $0.58\sigma_{\rm prior}$ for NFW and PF, respectively,  as shown in Fig. \ref{fig:3F_fits_radius} and summarized in Table \ref{tab:mass_shift}. The quantity $\Delta\mu/\sigma_{\rm prior}$ measures the shift of the 3F mean relative to the prior width, while $\sigma_{\rm 3F}/\sigma_{\rm prior}$ compares the dispersion of the 3F posterior medians across galaxies with the width of the corresponding prior Gaussian. In all three components, the dispersion of the posterior median masses across the galaxy sample is smaller than the width of the corresponding prior Gaussian. This is not a direct posterior contraction, since $\sigma_{\rm prior}$ is the width of the GMM-derived prior distribution, but it shows that the masses recovered in the 3F fits remain concentrated around the characteristic mass scales inferred from the 1F analysis.

The results from the three fermion analysis, presented in Figure \ref{fig:3F_fits_radius}, show a dependence on the radial extent of the rotation curves for the $m_{f,1}$ and $m_{f,2}$ parameters. As $R_{\max}$ increases, the posterior median masses tend to deviate more noticeably from the prior means toward lighter masses, as lighter masses produce larger halos. Another effect of the radial extent is a narrowing of the posterior intervals as $R_{\max}$ increases, suggesting that rotation curves extending to larger radii provide stronger constraints on the fermion masses. Still, despite these shifts, most of the inferred masses remain within the $2\sigma$ regions of the priors, indicating that the characteristic mass ranges obtained from the single-fermion analysis are broadly preserved.

\begin{table}[t]
\centering
\caption{Comparison between the characteristic mass scales obtained from the 1F GMM analysis and the distribution of posterior median masses from the 3F fits.}
\label{tab:mass_shift}
\begin{tabular}{cccccc}
\hline\hline
Halo & Mass & $\mu_{\rm prior}$ [eV] & $\mu_{\rm 3F}$ [eV] & $\Delta\mu/\sigma_{\rm prior}$ & $\sigma_{\rm 3F}/\sigma_{\rm prior}$ \\
\hline
NFW & $m_{f,1}$ & 21.52 & 19.92 & -0.24 & 0.83 \\
    & $m_{f,2}$ & 42.34 & 41.33 & -0.10 & 0.85 \\
    & $m_{f,3}$ & 71.26 & 77.23 &  0.46 & 0.82 \\
\hline
PF  & $m_{f,1}$ & 20.40 & 19.08 & -0.21 & 0.78 \\
    & $m_{f,2}$ & 39.78 & 38.81 & -0.11 & 0.93 \\
    & $m_{f,3}$ & 65.39 & 73.88 &  0.58 & 0.80 \\
\hline\hline
\end{tabular}
\end{table}

\section{model comparison and discussion}\label{sec:model-comparison}

The addition of parameters gives models more flexibility, which usually results in better description of the phenomena --maximizing the likelihood or minimizing the $\chi^2$-- but these added parameters might seem unjustified when a simpler alternative model is capable of similar results.

This principle is known as Occam's Razor and states: when two explanations provide the same outcomes, the less complex hypothesis should be preferred. 

Statistically speaking, this is interpreted as favoring the model with the fewest parameters, that provides an adequate description, as even if the results are better, it might just be a case of overfitting the data, instead of giving a faithful description of it.

Our dilemma is wether a physically motivated model, which emulates what we know of the universe might be regarded as \textit{simple}, even if it is mathematically more complex. In our case, this corresponds to taking a page out of nature's book and proposing a dark sector that replicates the generational structure of the Standard Model.

\begin{figure*}
    \centering
    \includegraphics[width=\linewidth]{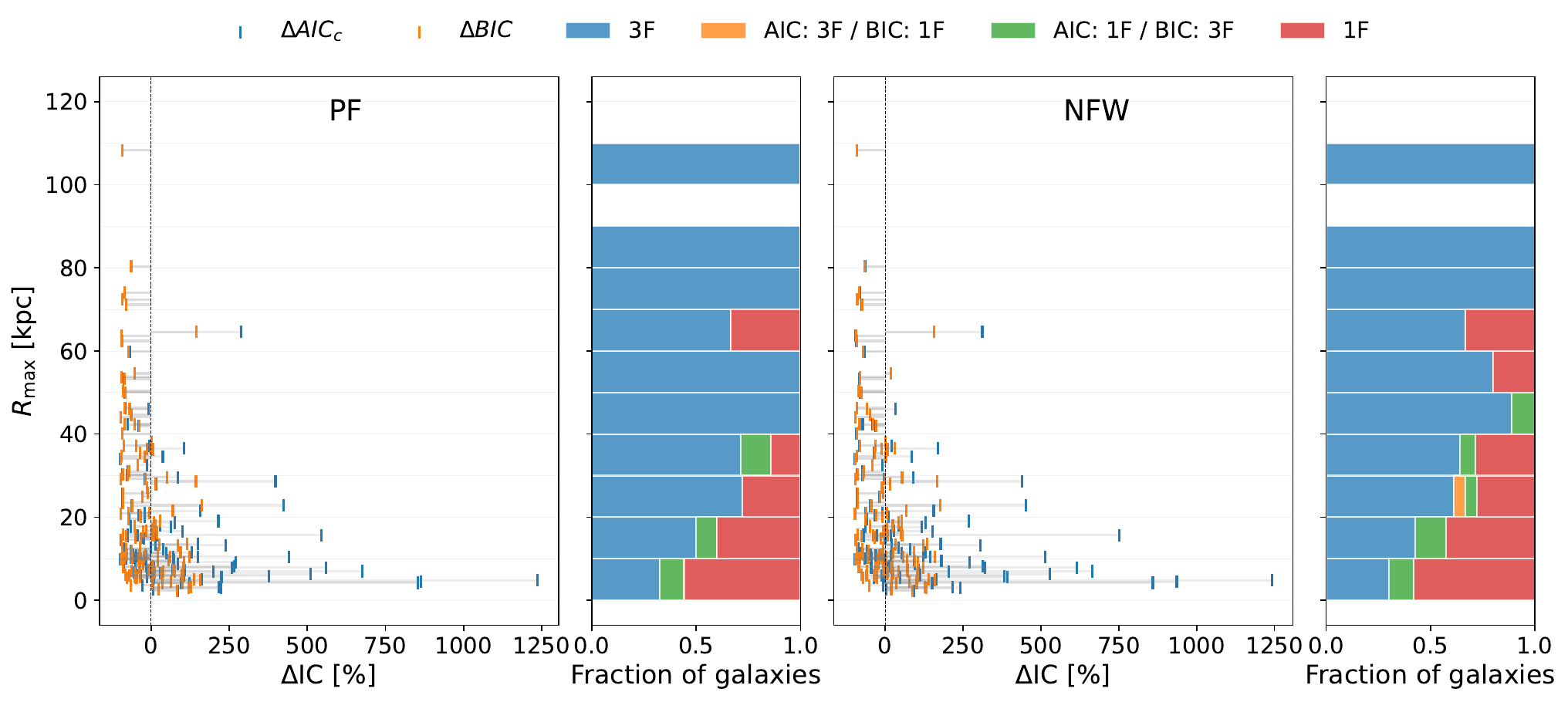}
 \caption{Comparison between the one- and three-fermion models using the AICc and BIC criteria. Left panels show the relative information-criterion differences as a function of the maximum observed radius $R_{\max}$. Right panels show the fraction of galaxies for which both criteria favor the 3F model, both favor the 1F model, or the two criteria give different preferences.}
    \label{fig:preference_fraction}
\end{figure*}

To evaluate the trade-off between complexity and goodness of fit, several model-selection criteria have been developed \cite{Liddle2007}, such as the Bayesian Information Criterion (BIC) \cite{Schwarz1978}

\begin{equation}
    BIC=k\ln{N}-2\ln{\mathcal{L}_{\max}}
\end{equation}

where $\cal{L}_{\max}$ is the maximum of the likelihood function and $k$ is the number of model parameters, penalizing models with a larger number of parameters. Another way to measure this cost-benefit is the  Corrected Akaike Information Criterion (AICc)\cite{Sugiura1978}, the finite-sample correction of the original AIC criterion \cite{Akaike1974}

\begin{equation}
    AICc=\frac{2kN}{N-k-1}-2\ln{\mathcal{L}_{\max}},
\end{equation}

this criterion includes the number of data points $N$ and reduces to the traditional AIC when $N>>k$.

The model which best manages the trade-off between complexity and goodness of fit has the smallest value of the information criterion value. It is convenient to evaluate the differences between the information criterion quantities to easily distinguish the relative preference between the models, or lack thereof, in the resulting values of AICc and BIC:

\begin{align}
    \Delta AICc &= AICc_{3F} - AICc_{1F},\\
    \Delta BIC &= BIC_{3F} - BIC_{1F}
\end{align}

Thus, $\Delta AICc <0$ or $\Delta BIC <0$ indicates a preference for the 3F model case to the corresponding information criterion, and $\Delta AICc >0$ or $\Delta BIC >0$ would mean that the improvement in reproducing the data, reflected in the maximum likelihood, is not enough to justify the added parameters.

It is important to note that the criteria might not agree, as they penalize the additional parameters differently.

From the 159 galaxies remaining after the 1F analysis, the restriction $N>7$ imposed by the six-parameter 3F model for the AICc comparison reduces the sample to 137 galaxies. The same sample of galaxies was used for the BIC analysis to maintain consistency. 

The additional degrees of freedom of the 3F model lead to an improvement in the minimum $\chi^2$ for 130 of the 137 galaxies (94.9\%) in both the NFW and PF cases. Across the two halo cases, there are 14 cases where we find no improvement, the relative increase in $\chi^2$ is below $5\%$ for 12 of the 14.

When we evaluate the information criteria, we can see clearly the penalty due to the additional parameters. In the NFW case, AICc favors the 3F model for 51.1\% of the galaxies and BIC for 60.6\%, while both criteria simultaneously favor the 3F model in 50.4\% of the sample. For the PF case, the 3F model is preferred in 56.9\% and 65.0\%, for the AICc and BIC methods, respectively, and in 56.9\% of the cases, it is preferred by both criteria. These results are summarized in Fig. \ref{fig:preference_fraction}, where the relative percentage differences of AICc and BIC are shown as a function of the maximum observed radius, $R_{\max}$. 

In this figure a trend is apparent: as $R_{\max}$ increases, the 3F model becomes more favored. This suggests that galaxies with larger $R_{\max}$ are more difficult to reproduce with the 1F model and the added parameters seem more justifiable.

\begin{figure*}
    \centering
    \includegraphics[width=\linewidth]{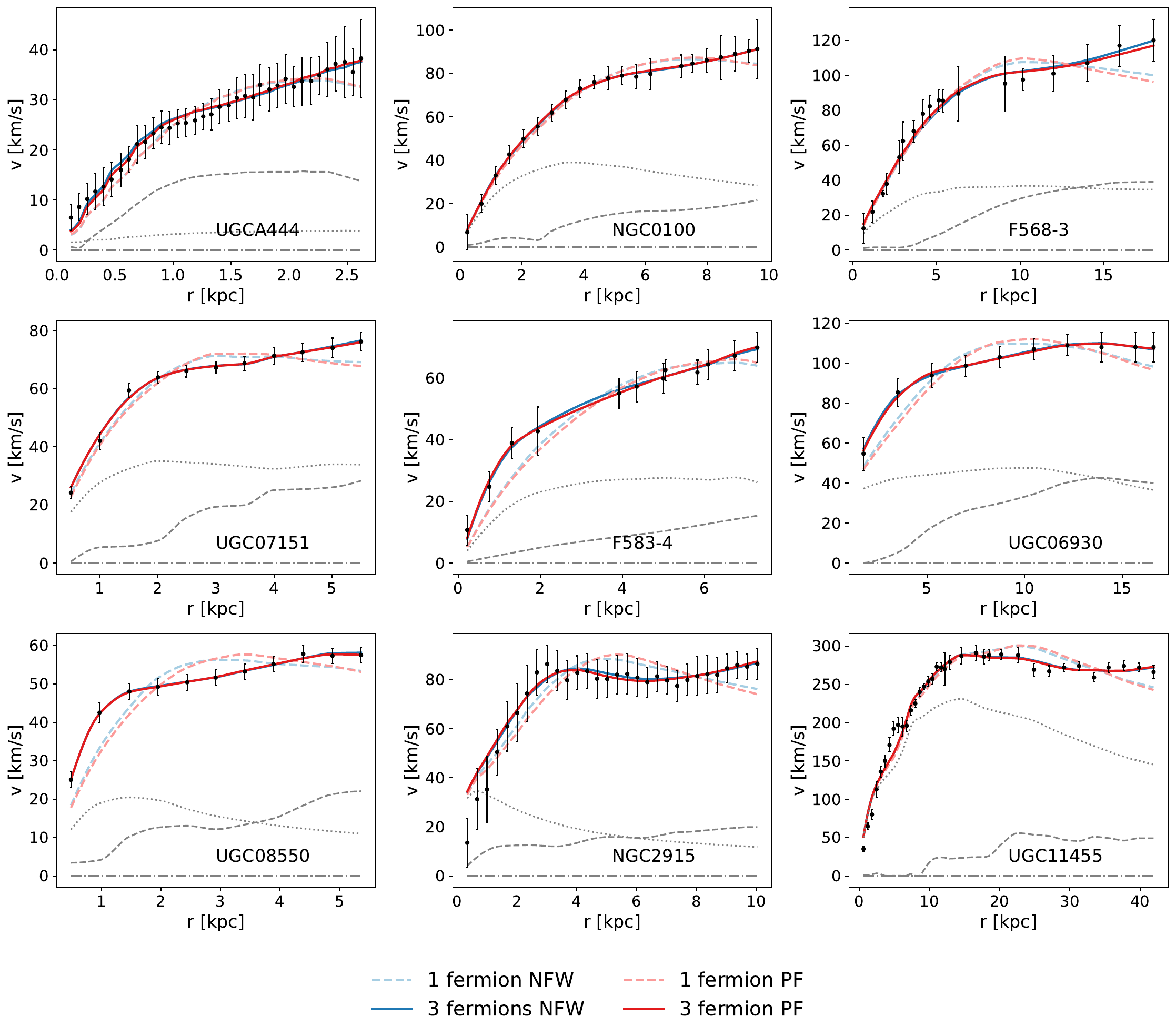}
    \caption{Representative SPARC rotation-curve fits for the one (dashed) and three-fermion (solid) models, with blue and red corresponding to the NFW and PF halos, respectively. The gray curves indicate the baryonic contributions to the total rotation curve. Galaxies are grouped according to the AICc and BIC model comparison: the upper row contains 1F-preferred cases, the middle row cases with mixed preference between the two criteria, and the lower row 3F-preferred cases.}
    \label{fig:example_rot_curve_3F}
\end{figure*}

Galaxies with simpler rotation-curve morphologies and smaller $R_{\max}$ can generally be described sufficiently well by the one-fermion model; this is evident in both the quality of the fits and the information criteria. The additional degrees of freedom of the three-fermion model become increasingly justified for galaxies with more complex rotation-curve structures, which in many cases are associated with larger values of $R_{\max}$. Representative examples are shown in Fig. \ref{fig:example_rot_curve_3F}. The upper row contains galaxies for which both AICc and BIC favor the 1F model, the middle row shows cases in which the two criteria give different preferences, and the lower row contains galaxies for which both criteria favor the 3F model. It should also be emphasized that the ability of the 1F model to reproduce a wide variety of rotation curves relies on keeping the fermion mass as a free parameter for each galaxy. If the dark matter is composed of a single fermionic species, however, its particle mass should be fundamental rather than galaxy dependent.

\section{Conclusions}\label{sec:conclusion}

Taking inspiration from the three generations of fermions in the Standard Model, in this work we proposed a dark matter sector composed of three fermionic species with hierarchical masses that only interact with the Standard Model gravitationally. 

Assuming that dark matter is composed of three fermions, we investigate wether at galactic scales the model can reproduce the observed rotation curves of galaxies. In order to do so, we model the dark matter halo as the self-gravitating configurations composed of three massive fermions.
First, we developed the self-gravitating system of fully degenerate fermions. The three components are gravitationally coupled and we found that the presence of the other species of fermions modifies the density profile relative to the corresponding single-fermion case. The components closer to the center appear to become more compact, while the outermost component becomes more extended, and diluting more rapidly. These changes are reflected in the rotation curve, which is not a simple superposition of the single-fermion systems. The multi-component system provides the capacity to describe more complex rotation curves.

We then establish an effective temperature for the system that determines the transition from a fully degenerate fermion gas to either a perfect-fluid-like halo or the standard CDM density profile of NFW. Each component has a particular Fermi temperature and they transition independently to the non-degenerate halo.
 
To identify physically motivated mass regions for the three fermions, we first analyzed the SPARC galaxy sample within the single-fermion scenario. The distribution of posterior median fermion masses was found to be multimodal. A three-component Gaussian mixture model identified the characteristic mass scales. For the perfect fluid case:  $m_{f,1} = 20.4\pm6.2$ eV, $m_{f,2} =39.8\pm8.6$ eV, $m_{f,3} =65.4\pm14.5$ eV. For the NFW halo, the Gaussian components are: $m_{f,1} =21.5\pm6.6$ eV, $m_{f,2} =42.3\pm9.9$ eV, $m_{f,3} =71.3\pm13.0$ eV. This implies that there is only a weak dependence on the assumed outer-halo model.

These distributions were used to construct data-informed Gaussian priors for the three masses in the 3F analysis. The results of the Bayesian analysis with the three-fermion model show that the mass ranges inferred from the 1F analysis remain largely preserved when the galaxies are fitted with the full three-fermion model, with the third component having the largest shift toward somewhat larger masses. The mean values obtained in the 3F analysis are $\mathbf{m}_f=(19.92,41.33,77.23)$ eV for the NFW case and $\mathbf{m}_f=(19.08,38.81,73.88)$ eV for the perfect-fluid case. The largest shifts, corresponding to the third component, amount to $0.46\sigma_{\rm prior}$ and $0.58\sigma_{\rm prior}$ for NFW and PF, respectively.  

For many of the shifts, there is slight dependence on the radial extent of the rotation curves, especially for the $m_{f,1}$ and $m_{f,2}$ parameters. As $R_{\max}$ increases, the posterior median masses show a larger offset from the prior means into lighter masses, because of the need to produce larger halos. These galaxies also present a narrowing of the posterior intervals, suggesting stronger constraints on the fermion masses. Most of the galaxies that shift from the prior mean still remain within the $2\sigma$ regions, indicating that the ranges are mostly preserved.

The additional degrees of freedom of the 3F model improve the goodness of fit for 130 of the 137 galaxies considered in the model comparison, corresponding to 94.9\% of the sample for each halo case. This is expected as the added free parameters make the model more flexible, and an improvement in goodness of fit alone does not establish whether the additional parameters are justified.

To give a sense of the cost-benefit of the added parameters in relation to the goodness of fit we computed and compared the BIC and AICc of both models, both of which penalize the extra parameters. Once model complexity is taken into account, AICc favors the 3F model for 51.1\% of the NFW cases and 56.9\% of the PF cases, while BIC favors it for 60.6\% and 65.0\%, respectively. Both criteria simultaneously favor the 3F model for 50.4\% of the NFW sample and 56.9\% of the PF sample.

The comparison also indicates that the relative performance of the models depends on the radial extent and structure of the observed rotation curve. Galaxies with smaller $R_{\max}$ and simpler rotation-curve morphologies can generally be reproduced adequately by the one-fermion model, whereas the additional freedom of the three-fermion configuration becomes increasingly useful for systems extending to larger radii and displaying more complex structures.    

An important distinction between the two scenarios is that the single-fermion model reproduces the diversity of SPARC rotation curves by allowing the fermion mass to vary independently from galaxy to galaxy. If dark matter corresponds to a fundamental fermionic species, however, its mass should not depend on the particular galaxy being modeled. The three-fermion scenario instead provides three characteristic mass ranges while allowing their relative density distributions to vary between galaxies. Within the present analysis, this offers a possible way of reconciling the diversity of galactic rotation curves with a particle interpretation based on a fixed set of fermionic species.

\subsection*{Acknowledgements}

F. Hernandez-Gutierrez acknowledges financial support from the Secretaría de Ciencia, Humanidades, Tecnología e Innovación (SECIHTI) through a graduate scholarship.


\bibliography{bibliography}
\clearpage
\end{document}